# Twitter Spam Detection:
# A Systematic Review

Sepideh Bazzaz Abkenar, Mostafa Haghi Kashani*, Mohammad Akbari, and Ebrahim Mahdipour

**Abstract**—Nowadays, with the rise of Internet access and mobile devices around the globe, more people are using social networks for collaboration and receiving real-time information. Twitter, the microblogging that is becoming a critical source of communication and news propagation, has grabbed the attention of spammers to distract users. So far, researchers have introduced various defense techniques to detect spams and combat spammer activities on Twitter. To overcome this problem, in recent years, many novel techniques have been offered by researchers, which have greatly enhanced the spam detection performance. Therefore, it raises a motivation to conduct a systematic review about different approaches of spam detection on Twitter. This review focuses on comparing the existing research techniques on Twitter spam detection systematically. Literature review analysis reveals that most of the existing methods rely on Machine Learning-based algorithms. Among these Machine Learning algorithms, the major differences are related to various feature selection methods. Hence, we propose a taxonomy based on different feature selection methods and analyses, namely content analysis, user analysis, tweet analysis, network analysis, and hybrid analysis. Then, we present numerical analyses and comparative studies on current approaches, coming up with open challenges that help researchers develop solutions in this topic.

——————————— ◆ ———————————

## 1 INTRODUCTION

SOCIAL network is a platform of individuals that can be represented as a graph in which nodes or points are its users, and the relations between them are shown as edges and lines of the graph. Inevitably, Online Social Networks (OSNs) are critical collaboration tools for users in their daily lives [1]. Users join social networks in order to communicate with their friends, family, or those in whom they are interested via sending messages, sharing photos, videos, posting their opinions, and broadcasting news [2]. The data analysts use social network data such as Twitter to extract users' mindsets and customers' opinions, reveal potential trends of markets, detect competitive intelligence, gauge the reaction of markets to product releases, and track customer complaints. So, the challenge of applying social network data is to distinguish fake users, spams, and spammers in order to take advantage of the social network's data [3].

As a significant microblogging platform, Twitter, created in 2006, has drawn the attraction of users by presenting free services to spread messages up to 280 characters all over the globe [4]. However, the ever-increasing popularity of Twitter and the convenience application of this platform is accompanied by spammers' activities. Spammers post unsolicited tweets, which usually contain trending hashtags, or malicious URLs to deceive and direct users to the sites including malicious posts to reach their personal gains such as phishing, scam, spams, spreading viruses, and so forth [5]. As a result, both Twitter and researchers employ various detection strategies to combat spammers. Twitter has provided different ways for users to report unwanted tweets that are suspected of being spams [6].

Suspicious accounts that usually send duplicate contents or the same contents to multiple users or post tweets that only include URL contents can be marked and reported as spammers by users [6], [7]. Moreover, Twitter itself applies blacklisting services such as Trend Micro, called Web Reputation Technology, for filtering spams [8], but because spammers are constantly changing their attack strategies, blacklists services have limitations and are unable to detect spams as early as possible. Thus, researchers have leveraged Machine Learning (ML) methods to identify the underlying patterns of spammers' activities [9].

ML-based detection methods include multiple steps. Since ML deals with the data world, uses data as an input, aims to derive the information from the raw data, and makes predictions, the first step is collecting data. One of the usual options to acquire Twitter data is through Twitter Application Programming Interface (API). However, Twitter API has some drawbacks, such as accessing current tweets and historical tweets. The last 3200 tweets and the tweets of the recent 7-9 days can be extracted through API, but due to Twitter's policy, researchers also are not allowed to publish the tweet itself [10]. Another

————————————————

- *Sepideh Bazzaz Abkenar and Ebrahim Mahdipour are with the Department of Computer Engineering, Science and Research Branch, Islamic Azad University, Tehran, Iran. E-mail: sepideh.bazzaz@srbiau.ac.ir, mahdipour@srbiau.ac.ir.*
- *Mostafa Haghi Kashani is with the Department of Computer Engineering, Shahr-e-Qods Branch, Islamic Azad University, Tehran, Iran. E-mail: mh.kashani@qodsiau.ac.ir.*
- *Mohammad Akbari is with the Department of Computer Science, Amirkabir University of Technology, Tehran, Iran. E-mail: akbari.ma@aut.ac.ir.*
- *This work has been submitted to the IEEE for possible publication. Copyright may be transferred without notice, after which this version may no longer be accessible.*

way to access Twitter's data is to find the datasets that have been collected and published earlier by other researchers to satisfy research objectives. The collected data should be preprocessed and normalized for eliminating duplicate and missing values, and in the case of biased datasets, they should be resampled. Then feature engineering is performed to extract the most efficient features from tweets in order to distinguish spam and non-spam [11]; a model is chosen to serve the researcher's goals.

Further, in supervised ML approaches, a subset of a dataset is labeled as spam or non-spam, and the selected model is learned by employing a labeled dataset, while in unsupervised ML approaches, the model is trained from a non-labeled dataset. After the training step, the model is evaluated by an unused and new dataset called testing data to investigate how the trained model can detect spam and non-spam in a new dataset. After the evaluation, maybe it is necessary to tune the parameters to have more accurate outcomes. Finally, the prediction questions can be answered.

Until today, with our observations and scrutinies, no comprehensive Systematic Literature Review (SLR) has been written on Twitter spam detection that complicates the identification and assessment of the existing spam detection methods, research challenges, and gaps precisely. To the best of our knowledge, one SLR [12] was conducted on papers between 2009 and 2013 , but evaluation parameters and tools were not examined. Moreover, due to the popularity of social networks and Twitter in particular, a research plan for Twitter spam detection is investigated. An SLR presents a comprehensive review of state-of-the-art to reveal existing methods, challenges, and potential future research directions for research communities [13], [14].

We conduct an SLR with the intention of *identifying, classifying, comparing Twitter spam detection methods, evaluating the methods of existing papers systematically, and offering a reasonable taxonomy*. Additionally, to attain this intension and find the answers to the following research questions, this methodological review is conducted:

- What are the main motivations behind spam detection in Twitter?
- What are the existing Twitter spam detection algorithms, approaches, and the main merits and demerits of each approach?
- What are the parameters generally employed in the performance evaluation of spam detection on Twitter?
- What are the available tools and evaluation techniques used in Twitter spam detection?
- What are the current open issues and future challenges in the scope of spam detection on Twitter?

We followed the guidelines in [13], [15], [16], and [17]. We intend to explore the available Twitter spam detection approaches systematically and categorize them taxonomically and present a precise comparison analysis of approaches and their potential challenges and limitations.

This SLR presents a systematic review of current studies on spam detection techniques and solutions on Twitter. For this purpose, 55 papers are chosen and compared to introduce a scientific taxonomy for the classification of Twitter spam detection. We summarize available methods, main ideas, applied tools, advantages, disadvantages, and evaluation parameters, and provide statistical and analytical reports on them. Furthermore, an abreast list of the primary challenges and open issues are outlined, and the significant areas are defined where future researches can improve the methods of the selected papers.

The main purpose of this paper is to explore and compare the Twitter spam detection methods thoroughly. This review identifies the motivation for presenting an SLR. The results of this study are beneficial for researchers to find out the challenges and potential future research directions.

The remainder of this SLR is structured as Fig. 1: In Section 2, the previous related works are reviewed to position the contributions of spam detection in Twitter. The paper selection process, research methodology, and research questions are presented in Section 3. Section 4 explains our presented taxonomy. Section 5 discusses and compares the analysis of the results. Section 6 describes open issues and future directions. Threats to validity are given in Section 7. Lastly, we conclude this SLR in Section 8. Additionally, Table 1 depicts a set of frequently used abbreviations in this SLR.

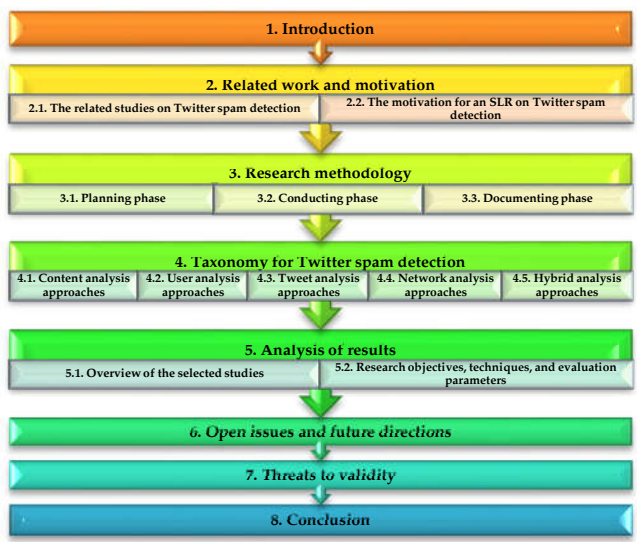

Fig. 1. The structure of this SLR

## 2 RELATED WORK AND MOTIVATION

To date, some reviews explore spam detection on Twitter, but literature reviews conducted in this subject have some limitations. In this section, we investigate the similarities and differences of current reviews on this topic according to the systematic research (Section 2.2), and the related works that drop in this topic are summarized in Section 2.1.

### 2.1 The Related Studies on Twitter Spam Detection

In this subsection, we intend to review the existing surveys on Twitter spam detection. Hence, several surveys have been explored below in this era. A comparison of surveys is represented in Table 2, in terms of several factors, including main ideas, publication years, open issues, paper selection processes, taxonomies, evaluation parameters, evaluation tools, and time range.

TABLE 1
LIST OF ABBREVIATIONS

| Abbr. | Full Form |
|---|---|
| OSN | Online Social Network |
| ML | Machine Learning |
| API | Application Programming Interface |
| SLR | Systematic Literature Review |
| SMS | Systematic Mapping Study |
| TF-IDF | Term Frequency- Inverse Document Frequency |
| KNN | K-Nearest-Neighbor |
| EM | Expectation-Maximization |
| CBOW | Continuous Bag-Of-Words |
| SG | Skip-Grams |
| SVM | Support Vector Machine |
| ELM | Extreme Learning Machine |
| HMM | Hidden Markov Model |
| CSC | Cost-Sensitive Classifier |
| HMPS | Hierarchical Meta-Path Scores |
| DT | Decision Tree |
| NN | Neural Network |
| NB | Naive Bayesian |
| RF | Random Forest |
| GA | Genetic Algorithm |
| PSO | Particle Swarm Optimization |
| PCA | Principal Component Analysis |
| LFUN | Learning From Unlabeled |
| ASL | Asymmetric self-learning |
| FOS | Fuzzy Oversampling |
| INB | Incremental Naïve Bayes |
| MLP | Multi-Layer Perceptron |
| GSB | Google Safe Browsing |
| FPR | False Positive Rate |
| FNR | False Negative Rate |

Wu, et al. [18] introduced a survey that offered a taxonomy for spam detection techniques on Twitter. The authors explored the papers in this scope and classified spam detection techniques in three categories, namely syntax analysis, feature analysis, and blacklist. In this research, the advantages and disadvantages of each method were mentioned, as well as their evaluation parameters and unsolved challenges. However, they did not present a systematic review, and the paper selection process, the time range of selected paper, and the evaluation tools were not specified. Further, Kabakus and Kara [19] surveyed the relevant papers about spam detection on Twitter. They also classified features that assist in spam detection in three categories: account-based, tweet-based, and graph-based features. Moreover, a taxonomy for spam detection was provided, as well as the pros and cons of each approach. The paper was not written in a systematic manner and the evaluation parameters, tools and the time range of papers were not mentioned.

In another research, Li and Liu [20] conducted a comparative study to explore the most effective resampling method for handling the imbalanced datasets. Further, an extensive comparative study was conducted to investigate the most robust classification algorithm to detect Twitter spams. The evaluation results in terms of accuracy, recall, Kappa, and F-measure revealed that the fuzzy-based ensemble technique outperformed in both resampling and classification compared to other techniques. Since a certain taxonomy was not proposed, this paper did not present an SLR. Moreover, Gheewala and Patel [21] carried out a survey to explore spam detection in the Twitter dataset. The authors categorized the detection scheme into classification, clustering, and hybrid approaches. They examined the extracted features, classification algorithms, and the results of the reviewed papers, but the paper selection process, evaluation parameters, applied tools, and open issues were not explored.

In another study, Lalitha, et al. [22] reviewed papers on Twitter spam detection published from 2010 to 2017. The detection approach was classified into user-based, content-based, and hybrid techniques. The pros and cons of each category, evaluation parameters, and unsolved challenges were studied, but applied tools and paper selection processes were not specified. Additionally, Daffa, et al. [23] provided a survey on state-of-the-art in spam URLs detection topic and explored spam behavior and feature classification as well as ML classification performance. The authors classified the features employed in spam URLs detection in three categories: user-based, content-based, and hybrid-based features. Findings indicated that applying hybrid-based features along with RF classifier had the highest performance in spam detection. However, paper the selection processes, open issues, and the time range of the selected papers were not indicated.

In addition, Imam and Vassilakis [24] provided a taxonomy of attacks against spam detectors in four categories including causative integrity, causative availability, exploratory integrity, and integrity availability, that can be employed by an adversary to deceive the trained model. Although the defense and implementation approaches in Twitter spam detection were offered, but the paper selection processes, evaluation metrics, and tools were not declared. Further, other Twitter spam detection surveys were perused such as [25] by surveying various approaches to diagnose spam accounts, [26] by reviewing spam detection papers published from 2010 to 2015, [12] by reviewing existing techniques for spammer detection, [27] and by introducing the definition of spam and different kinds of spam in educators' Twitter use. Additionally, the weak and strong aspects of the spam detection approaches were evaluated as well as the metrics that assisted educators in identifying and reacting to spams.

Spirin and Han [28] have conducted a survey on web-spam definition, various types of spams, along with web spam detection algorithms. In [29], researchers have reviewed current approaches for spam detection in social networks. In another survey, Kaur, et al. [30] offered two taxonomies for both spam detection techniques and compromised account detection approaches in social networks. The authors thoroughly reviewed the relevant papers qualitatively and investigated their pros and cons. Although they revealed the existing gaps for further research by a comparative analysis of the existing spam detection techniques, the survey did not perform systematically.

While Soman and Murugappan [31] explored some of the spam detection algorithms on social networks, Dhaka and Mehrotra [32] discussed the cross-domain spam detection techniques in social networks and their related challenges. Similarly, Krithiga and Ilavarasan [33] surveyed spam profile detection methods in social networks. Other researchers in [34] conducted an SMS to examine the spam detection methods on social networks and their architectural framework and outlined the limitation of the existing research.

TABLE 2
SUMMARY OF THE RELATED WORKS

| Type | Ref | Main idea | Pub. year | Open issue | Paper selection process | Taxonomy | Evaluation parameter | Evaluation tool | Time range |
|---|---|---|---|---|---|---|---|---|---|
| Survey | [18] | Twitter spam detection | 2018 | Clear | Not clear | Yes | Presented | Not presented | Not mentioned |
| | [19] | Twitter spam detection features | 2017 | Clear | Not clear | Yes | Not presented | Not presented | Not mentioned |
| | [20] | Class imbalance problem in spam detection | 2018 | Not clear | Not clear | No | Presented | Not presented | Not mentioned |
| | [21] | Twitter spam detection based on ML | 2018 | Not clear | Not clear | Yes | Not presented | Not presented | Not mentioned |
| | [22] | Twitter spam detection techniques | 2017 | Clear | Not clear | Yes | Presented | Not presented | 2010-2017 |
| | [23] | Twitter spam URLs detection | 2018 | Not clear | Not clear | Yes | Presented | Not presented | Not mentioned |
| | [24] | Attacks against Twitter spam detectors | 2019 | Clear | Not clear | Yes | Not presented | Not presented | Not mentioned |
| | [25] | Detecting spam accounts in Twitter | 2019 | Not clear | Not clear | No | Presented | Presented | Not mentioned |
| | [26] | Spam detection on Twitter | 2016 | Not clear | Not clear | Yes | Presented | Not presented | 2010-2015 |
| | [27] | Detecting spam and educators' Twitter use | 2019 | Clear | Not clear | No | Presented | Not presented | Not mentioned |
| | [28] | Web spam detection methods | 2012 | Clear | Not clear | Yes | Not presented | Not presented | Not mentioned |
| | [29] | Examing spam detection in social networks | 2015 | Not clear | Not clear | Yes | Presented | Not presented | Not mentioned |
| | [30] | Social network spam and compromised account detection | 2018 | Clear | Not clear | Yes | Not presented | Not presented | Not mentioned |
| | [31] | Spam detection algorithms on social networks | 2014 | Not clear | Not clear | No | Not presented | Not presented | Not mentioned |
| | [32] | Cross-domain spam detection in social networks | 2019 | Clear | Not clear | Yes | Not presented | Not presented | Not mentioned |
| | [33] | Spam profile detection methods in social networks | 2019 | Clear | Not clear | No | Presented | Not presented | Not mentioned |
| SMS | [34] | Investigating spam detection on social networks | 2017 | Clear | Clear | No | Not presented | Not presented | 2002-2016 |
| SLR | [12] | Spammer detection techniques in Twitter | 2014 | Clear | Clear | Yes | Not presented | Not presented | 2009-2013 |
| | This Study | Investigating spam detection approaches on Twitter | -- | Clear | Clear | Yes | Presented | Presented | 2010-April 2020 |

## 2.2 The Motivation for an SLR on Twitter Spam Detection

The need for an SLR is to identify, classify, and compare the existing research reviews in Twitter spam detection. In order to show that a similar SLR has not already been proposed, we searched Google Scholar (on April 2020) with the following search string:

(Twitter <OR> " social media" <OR>" social networks" <OR>" social network")
[AND]
Spam [AND]
(Review <OR> Survey <OR> Overview <OR> Challenges <OR> "open issues"
<OR> Trends <OR> literature <OR> Study <OR> "state-of-the-art")

As shown in Table 2, most of the retrieved reviews on Twitter spam detection were not conducted in a systematic manner, and their paper selection processes was unclear. Some studies have not proposed any lucid classification on this topic. Moreover, some papers have not declared the time ranges, evaluation parameters, and used tools in their reviews. Others have not declared open issues accurately, and future challenges were mentioned briefly. So, considering the importance of Twitter spam detection, highlighting open issues and feature research directions precisely is timely.

## 3 RESEARCH METHODOLOGY

To make a lucid and precise background of Twitter spam detection, this section presents an SLR method. An SLR is a methodology to investigate a specific subject that was raised in medical fields for the first time [35], [36] and now can be applied in any field of study. In contrast to a non-systematic review processes, an SLR is a process of performing taxonomical review and following a precise methodological analysis of the research literature to answer a specific research problem, which reduces the bias in a field of study. Since most review articles on Twitter spam detection were written in a non-systematic and unstructured manner, the purpose of this paper is to provide a rigorous process of methodological steps for researching literature in the scope of analyzing spam detection approaches on the Twitter social network.

In this systematic process, a three-phase guideline, including *planning*, *conducting*, and *documenting*, is adopted as depicted in Fig. 2 [13]. The review is accompanied by an external evaluation of the outcome of each phase. We first identify the questions and the needs that are the motivations of this SLR in the planning phase. Then the articles in this subject are extracted and selected based on inclusion/exclusion criteria in the conducting phase. Ultimately, in the documenting phase, the observations are documented, and the results of papers were analyzed, compared, and documented, yielding the answers to the research questions, then the final reports are presented. The three phases of research methodology followed in this paper are discussed below:

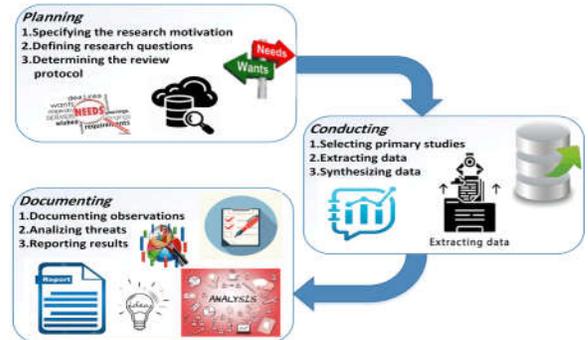

Fig. 2. Overview of research methodology

### 3.1 Planning Phase

Planning begins with an identification of the research motivation for this SLR and finishes in a review protocol as follows:

**Stage 1-** *Specifying the research motivation.* According to the contribution of this SLR that is justified by comparing the available surveys explained in Section 2.2, the motivation is specified at the first stage.

**Stage 2-** *Defining research questions.* As mentioned earlier, in the second stage, the research questions based on the motivation of this paper are defined in Table 3 that assist in developing and validating the review protocol. The research questions mentioned above are shown in Table 3.

**Stage 3-** *Determining the review protocol.* According to the goals of this SLR, in the previous stage, the research questions and the review scope were identified to adjust search strings for literature extraction [13]. Moreover, a protocol was developed by following [37] and our previous experience with SLR [35], [36], [38]. In order to evaluate the defined protocol before its execution, we requested an external specialist for feedback, who was experienced in conducting SLRs in this era. His feedback was applied in the upgraded protocol. A pilot study (approximately 20%) of the included papers was performed to reduce the bias between researchers and enhance the data extraction process. We also enhanced the review scope, search strategies, and inclusion/exclusion during the pilot stage.

TABLE 3
RESEARCH QUESTIONS

| |
|---|
| RQ1-What are the main motivations behind spam detection in Twitter? |
| RQ2-What are the existing Twitter spam detection algorithms, approaches, and the main merits and demerits of each approach? |
| RQ3-What are the parameters generally employed in the performance evaluation of spam detection on Twitter? |
| RQ4-What are the available tools and evaluation techniques used in Twitter spam detection? |
| RQ5-What are the current open issues and future challenges in the scope of spam detection on Twitter? |

## 3.2 Conducting Phase

The second phase of the research methodology is conducting, which starts with paper selection and concludes in data extraction. The aim of this subsection is to represent the process of searching and selecting papers conducted in the second phase of SLR. In order to choose papers in this era, we follow a two-step guideline, as depicted in Fig. 3.

- *Initial selection.* The following search strings are explored among academic databases to find papers that have these strings in their titles, abstracts, and keywords. In such a case, popular online academic databases such as Springer, IEEE Explorer, ScienceDirect, Taylor&Francis, Wiley, Google Scholar, and ACM are employed. **356** preliminary papers extracted at the initial selection are depicted in Table 4 based on each digital library. Moreover, we consider the papers published online between 2010 to 2020, in this step.

**(Twitter <OR> tweet <OR> tweets)**
**[AND]**
**(spam <OR> spammer <OR> malicious)**

- *Final selection.* The extracted 356 papers from the previous step were examined, and the inclusion/exclusion criteria were applied (as shown in Table 5). Non-English and survey papers, theses, books, book chapters, and short papers were excluded, then papers were investigated in full text, and quality assessment was applied, and only papers mentioning the evaluation details and their techniques explicitly were selected. As a result, **55** relevant studies were chosen at the end of this step to be assessed qualitatively.

TABLE 4
NUMBER OF TOTAL PAPERS

| No | Academic databases | Results |
|---|---|---|
| 1 | Springer | 52 |
| 2 | IEEE | 89 |
| 3 | ScienceDirect | 54 |
| 4 | Taylor&Francis | 3 |
| 5 | Wiley | 5 |
| 6 | Google Scholar | 120 |
| 7 | ACM | 33 |
| | *Total* | **356** |

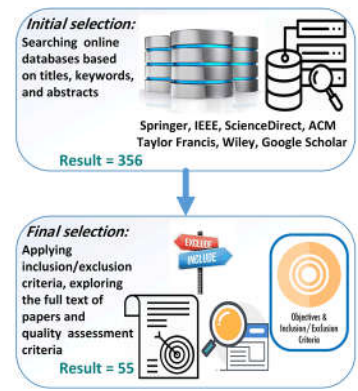

Fig. 3. The paper selection process

TABLE 5
INCLUSION/ EXCLUSION CRITERIA

| Inclusion criteria | Exclusion criteria |
|---|---|
| Journal, conference and peer-reviewed papers | Editorial papers, white papers, non-English papers and papers less than six pages |
| Studies that focus on spam detection on Twitter | Books, book chapters, theses and review papers |
| Research papers that present techniques or innovative solutions to enhance spam detection in the Twitter dataset | Research papers that do not mention solutions and methods to improve spam detection rate on Twitter dataset explicitly |

## 3.3 Documenting Phase

The papers were extracted from a list of seven popular databases (as shown in Table 4), then we proposed a classification on Twitter spam detection in Section 4. Further, the pros and cons of the existing papers and their challenges were analyzed in this phase.

## 4 TAXONOMY FOR TWITTER SPAM DETECTION

Since various spam detection methods may apply different techniques to tackle distinct aspects of spam detection, proposing a taxonomy of the current methods in Twitter spam detection is a difficult and challenging task. According to the state-of-the-art review, most Twitter spam detection techniques are based on ML methods that use various classification and clustering algorithms. Nonetheless, most of the applied ML algorithms are somewhat similar, and the main difference among these methods is in their feature selection strategies.

Thus, in this SLR, we adopt a taxonomy of the current methods in Twitter spam detection based on feature selection, as illustrated in Fig. 4. According to various feature selection extraction and strategies, we present five significant categories: *Content analysis approaches* [39], [40], [41], [42], [43],[44], [45], [46], *User analysis approaches* [47], [9], [48], [49], [50] *tweet analysis approaches* [51], [52], [53], [54], [55], *network analysis approaches* [56], [57], [58], [59], [60], [61] and *hybrid analysis approaches*. *Hybrid analysis approaches* combines *various features to accomplish spam detection, such as user and tweet analysis approaches* [11], [62], [63], [64], [65], [66], [67], [68], [69], [70], [71], [72], [73], [74], [75], *content, user, and tweet analysis approaches* [76], [7], [77], [78], *content and network analysis approaches* [79], [80], [81], [82], [83], *user, tweet, and network analysis approaches* [84], [85], [86], [87], *network, tweet, and content analysis approaches* [88], [89], or *network, content, and user analysis approaches* [90].

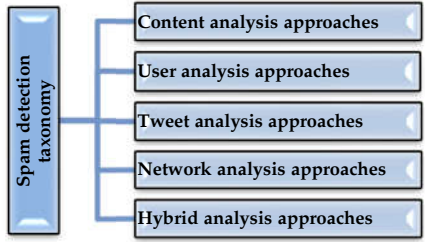

Fig. 4. Taxonomy of Twitter spam detection

## 4.1 Content Analysis Approaches

To begin with, we review the detection methods based on content analysis approaches. This approach detects spams by focusing on tweet contents, in which text analysis techniques are applied to discriminate between ham and spam contents. Since spammers usually use malicious content to spread spams, a tweet containing duplicate posts, or malicious words is likely to be spam. On the other hand, as Twitter posts have length limitations, spammers usually apply malicious HTTP links or keywords. So, in content analysis approaches, researchers discover spams by exploring the content of tweets via such techniques as a bag of words and Term-Frequency-based (TF-IDF). These techniques usually represent each document by a vector of term frequencies (TF) that demonstrates the ratio of occurrence of a particular word to the total number of words in the document. Inverse Document Frequency (IDF), on the other hand, measures the occurrence of a term in different documents of a corpus. The multiplication of TF and IDF constructs (TF-IDF) feature vectors for each word of a document and shows the relative importance of a term in a document as compared to other words in the corpus. Studies that are based on content analysis approaches are listed below:

Since blacklists classify URLs or IP addresses, spammers produce new short URLs as soon as the previous URLs are blacklisted, so, Wang, et al. [39] focused on short URLs and their click traffic features to distinguish Twitter spams. Santos, et al. [40] developed a compression-based text classification method to detect spam and ham tweets. In [41], the authors presented a spam detection system to find spam tweets by using content analysis approaches.

In [42], the authors offered a decentralized and distributed method that used content-based features along with a fuzzy string matching algorithm to detect and combat Twitter spams. Ashour, et al. [43] focused on exploring the impact of applying character n-gram features in real-time spam detection. In order to create the HSpam14 dataset, Sedhai and Sun [44] used heuristic-based methods to gather English tweets according to hashtags and popular keywords that were suspicious to be spam. Since spammers post their tweets multiple times with the least changes to be more visible, they applied the MinHash based algorithm. In this phase, both KNN-based and manual annotation were applied as well as the multiple tweets sent by the same users, and those containing tweets from similar domains. Positive tweets were utilized for the remained unbalanced tweets in the previous phase to detect ham tweets. Up to this phase, 7 million tweets were labeled out of 14 million collected tweets. At last, Expectation-Maximization algorithm (EM) was applied to label the remained unbalanced tweets.

Al-Azani and El-Alfy [45] proposed a spam detection method to detect spam posts by focusing on contents and employing Continuous Bag-Of-Words (CBOW) and Skip-Grams (SG) on Twitter Arabic contents. Additionally, [46] drop in this category. Table 6 depicts the main ideas, merits, demerits, and evaluation parameters of the studied papers, which are based on content analysis approaches.

## 4.2 User Analysis Approaches

In this section, the spam detection method based on user analysis approaches is discussed. In this method, user account statistic features are analyzed to detect spams. In other words, statistic features of user account profiles such as account_age, no_follower, and no_following are investigated. Findings show that spammers follow a large number of users in a short time, or they have a small number of followers in comparison with the number of followings. The studied papers that are based on user analysis approaches are summarized below:

Since some Twitter features are historical and available for a limited time, Inuwa-Dutse, et al. [47] proposed an approach that employed a set of account-based features to detect spam messages in real-time. Benevenuto, et al. [9] employed the SVM classifier and explored the impact of various features for spammer detection on Twitter. They examined the effect of the various subset of features, namely user-based features and a combination of user and tweet-based features that yield high performance in both subsets. Additionally, researchers in [48] employed user account features to detect spammer profiles.

Soman and Murugappan [49] extracted various types of features in which applying user-based features in trending topics along with Extreme Learning Machine (ELM) had the highest F-measure in Twitter spam detection in comparison with other types of features and classifiers. In order to detect malicious accounts, Venkatesh, et al. [50] focused on various user-based features along with a new feature calling trust score, which led to high accuracy.

Table 7 Depicts the main ideas, merits, demerits, and evaluation parameters of the studied papers, which fall in user analysis approaches.

TABLE 6
REVIEWING AND COMPARING CONTENT ANALYSIS APPROACHES

| Ref. | Main ideas | Merits | Demerits | Evaluation Parameters |
|---|---|---|---|---|
| [39] | Detecting spam by short URL analysis | • High scalability | • Not tested on various features | Accuracy, precision, recall, F-measure, scalability |
| [40] | Filtering spam based on contents of tweets | • High accuracy of RF classifier | • Not studying the linguistic relationships in tweets | ROC (AUC), accuracy, precision, recall, F-measure, |
| [41] | Filtering bi-lingual spam tweets by applying ML | • High accuracy | • Low scalability | Time, ROC (AUC), accuracy |
| [42] | Detecting spam by using fuzzy string matching algorithm | • High scalability | • High running time | Precision, recall, specificity |
| [43] | Detecting spam tweets by employing N-gram features | • High speed (low response time) | • Low robust<br>• Not employing user feedback and sending it into the system as new labeled data | Time, precision, recall, F-measure |
| [44] | Detection spam posts on hashtags of tweets | • Providing a large dataset for content-based spam detection for researchers | • Not accessing to full user profiles<br>• Not being able to perform user-level spam detection on this dataset | Jaccard similarity coefficient, cosine similarity, precision |
| [45] | Arabic tweet spam detection by using ML and embedded word | • High F-measure | • Low dimensionality | Accuracy, precision, recall, F-measure |
| [46] | An ontology-based spam detection method | • High accuracy<br>• High detection rate<br>• High scalability | • High running time | Accuracy, FPR, FNR |

TABLE 7
REVIEWING AND COMPARING USER ANALYSIS APPROACHES

| Ref. | Main ideas | Merits | Demerits | Evaluation Parameters |
|---|---|---|---|---|
| [47] | Proposing a real-time spam account detection | • High robustness<br>• High efficiency of real-time detection<br>• High F-measure on RF classifier | • Not testing the impact of the recent increase in the maximum length of tweets on spammer activities | Accuracy, Precision, Recall, F-measure |
| [9] | Presenting spammer detection techniques based on classification | • High accuracy<br>• High scalability | • Not tested on various classifiers | Precision, recall, F-measure |
| [48] | Applying labeling for Twitter spam account detection | • High accuracy<br>• High F-measure | • Only analyzing English tweets<br>• High execution time | Jaccard similarity coefficient, accuracy, F-measure, maximum cluster diameter |
| [49] | Detecting malicious tweets by clustering and classification techniques | • High F-measure | • High response time | Accuracy, F-measure, FPR, recall |
| [50] | Detecting malicious accounts based on Twitter short URLs | • High F-measure | • Low scalability | Accuracy, F-measure, FPR, recall |

TABLE 8
REVIEWING AND COMPARING TWEET ANALYSIS APPROACHES

| Ref. | Main ideas | Merits | Demerits | Evaluation Parameters |
|---|---|---|---|---|
| [51] | Applying a Markov model for real-time spam detection | • High recall | • Low precision<br>• Not applying the higher orders of HMM for real-time detection | Precision, recall, F-measure, |
| [52] | Proposing a hybrid model for spam tweet detection based on unsupervised language model | • High scalability<br>• Low response time | • Not testing on user accounts features<br>• Low F-measure | Accuracy, precision, recall, F-measure |
| [53] | Proposing a cost-sensitive classifier for Twitter spam detection | • High recall<br>• High accuracy | • Low scalability | Accuracy, recall, specificity |
| [54] | Detecting Twitter spam trending topics based on naive bayes classifier | • High F-measure | • Low scalability | Precision, recall, F-measure |
| [55] | Employing statistical analysis for malicious spam topics detection | • High recall | • Not testing the impact of unbalanced dataset<br>• High cost | Accuracy, recall, F-measure |

## 4.3 Tweet Analysis Approaches

In this section, papers that applied tweet analysis approaches are reviewed. This approach extracts statistic features of tweets such as no_tweets, no_retweets, no_hashtag, no_digits, and no_char to detect spams.

Washha, et al. [51] applied the first-order Hidden Markov Model (HMM) as a time-dependent prediction model and leveraged tweet-based features to identify spam posts in real-time. Washha, et al. [52] investigated how the collaboration of Twitter with Facebook could improve diagnosing spam posts of trending subjects in large datasets. Tur and Homsi [53] introduced a novel spam detection approach employing CSC and tweet-based features to increase the classification accuracy in an imbalanced dataset.

Additionally, [54] extracted the tweet features in the trending topics on Twitter and applied NB to detect spams and to explore whether those topics were misused by spammers. Instead of focusing on spam account detection, Martinez-Romo and Araujo [55] developed a language model by employing statistical tweet analysis to detect spam posts of trending topics on Twitter.

Table 8 illustrates the main ideas, merits, demerits, and evaluation parameters of the studied papers, which are based on tweet analysis approaches.

## 4.4 Network Analysis Approaches

As we mentioned earlier, the relations among users in social networks such as Twitter can be presented as graphs. In network analysis approaches, the communication graphs of Twitter are analyzed for spam detection. Thus, network analysis approaches focus on such attributes as distance and connectivity. The length of the distance and

TABLE 9
REVIEWING AND COMPARING NETWORK ANALYSIS APPROACHES

| Ref. | Main ideas | Merits | Demerits | Evaluation Parameters |
|---|---|---|---|---|
| [56] | Applying the relationship of sender-receiver for spam filtering | • High accuracy | • High cost (High bandwidth computing, storage resource and time) | ROC (AUC), accuracy, FPR |
| [57] | Employing retweet-relation for Twitter spam detection | • High accuracy | • Low precision and low recall in a high time interval | Transitivity, degree variance, local clustering coefficient, accuracy, precision, recall, F-measure |
| [58] | Detecting spam campaigns on Twitter | • High detection rate | • Low scalability | ROC (AUC), accuracy, precision, recall, F-measure |
| [59] | Detecting Twitter spammer based on graph-based and neighbor-based features | • High F-measure | • Low scalability | Local clustering coefficient, F-measure, FPR, detection rate, betweenness centrality |
| [60] | Proposing a trust model for detecting social botnet | • High F-measure | • Low scalability | Precision, recall, F-measure |
| [61] | Presenting a hybrid approach for malicious profile detection | • Better accuracy in terms of TPR and FPR<br>• Real-time fake account detection<br>• High detection rate | • High cost | ROC (AUC), accuracy, precision, recall, F-measure, FNR |

the strength of the relationship between accounts are examined. The papers that applied network analysis approaches in their presented method are listed in this subsection:

Song, et al. [56] presented a spam detection system to distinguish spammers in real-time. They used a directed graph that represented the relations between Twitter users and measured distance and connectivity features between the sender and receiver of messages. Chen, et al. [57] offered a novel detection system as a message-passing graph analyzer by extracting graph-based features including degree variance, clustering coefficient, and transitivity in which clustering and classification algorithms were run by setting 5 minutes for the unit time interval.

Gupta, et al. [58] developed a collective classification and feedback-based approach along with a Hierarchical Meta-Path Scores (HMPS) to measure the similarity between nodes in a heterogeneous network and to identify suspended accounts as spammers per campaign that was superior to previous spam detection and oversampling methods. Similarly, [59] drops in this category.

Lingam, et al. [60] designed a novel algorithm for diagnosing social botnet through applying a trust model. In [61], researchers offered a framework to distinguish Sybil and legitimate accounts by using a graph Petri net solution with various ML algorithms.

Table 9 shows the main ideas, merits, demerits, and evaluation parameters of the studied papers, which are based on network analysis approaches.

## 4.5 Hybrid Analysis Approaches

Hybrid analysis approaches use a combination of content analysis, user analysis, tweet analysis, or network analysis approaches to distinguish spam accounts from non-spam ones. The papers with hybrid approaches are summarized in this subsection:

Wang [79] applied content-based features along with graph-based features to diagnose spams from benign tweets. The author developed a web crawler through Twitter API, and four classification algorithms, including DT, NN, SVM, and NB, were employed. Findings indicated that NB outperformed others in terms of F-measure.

Mccord and Chuah [72] used a combination of content and user-based features such as the number of followers, followings, URLs, reputation, replies, mentions, hashtags, and retweets to simplify spam detection system. They evaluated the performance of suggested features by four classifiers. Findings showed that RF had high results in both precision and F-measure.

Chen, et al. [62] developed a system to improve spam detection by using feature discretization. They applied six ML algorithms to investigate the effect of feature discretization on continuous balanced and imbalanced datasets. As the performance of the classifiers reduces over time due to the changing of statistical features of spam tweets, Chen, et al. [63] offered the LFUN technique (Learning From Unlabeled tweets) to enhance the classifier algorithms performance. The proposed approach could identify "changed" spam tweets from unlabeled tweets to employ them in the process of training the model.

Lin and Huang [64] utilized two effective features, the URL rate and the interaction rate to distinguish spam from non-spam tweets. The detection rate was high in both precision and recall in this hybrid approach. Zhang, et al. [80] introduced a URL-based approach to distinguish legitimate campaigns from non-legitimate ones (spam or promoting), then by constructing social graphs and identifying dense subgraphs and applying SVM algorithm the spam and promoting campaigns were detected.

Chen, et al. [65] applied "changed spam" tweets in training the ML algorithm as an Asymmetric self-learning scheme (ASL) for the "spam drift" problem to reduce the performance degradation of the classifier. In order to solve the negative impact of the imbalanced dataset on the performance of classifiers, Liu, et al. [11] presented a fuzzy oversampling method (FOS) to produce synthetic data samples, and they also applied an ensemble method with a voting scheme to enhance the efficiency of classifiers.

Guo and Chen [88] applied four classifiers with employing content-based, graph-based, and geographic features to detect spam accounts. RF outperformed other classifiers in terms of F-measure, precision, recall, and root mean square error; however, DT predicted more non-spam ac-counts. Since spam tweets are drifting over time in the statistical feature scope, the performance of classifiers reduced. Thus, in order to address the Twitter spam drift problem, Liu, et al. [70] introduced fuzzy-based and asymmetric sampling techniques. A novel fuzzy-based method was employed to produce more spam instances and alleviate the unbalance distribution in an imbalanced dataset. Then an asymmetric method was developed to re-balance the sizes of spam and non-spam samples in training dataset. At last, the ensemble technique was applied to combine the classifiers and increase the accuracy of detection.

Miller, et al. [76] presented two modified clustering algorithms, namely StreamKM++ and DenStream, to detect spams in streaming tweets. Instead of solving spam detection as a classification problem, the authors applied a combination of two proposed algorithms to solve it as an anomaly detection method. A clustering model was built on normal users, and all outliers were considered as spams. Chu, et al. [89] proposed a spam detection framework by developing a clustering algorithm that clustered tweets according to shared URLs. Then the authors applied a combination of content and tweet features to train the RF classifier and distinguish spam tweets from legitimate ones.

Further, Sedhai and Sun [77] presented a semi-supervised method with two main modules, spam detection, and update module to diagnose spam and ham tweets. Nilizadeh, et al. [82] proposed a new system that applied a combination of various techniques including network science, natural language processing, and ML to detect spam tweets through communities among users with the same interests. Bindu, et al. [90] offered an unsupervised method that used context, account, and graph-based features to detect spammer communities. Adewole, et al. [87] developed K-means algorithms and used PCA to address the problem of enhancing initial recognition of spammer accounts, further, three classifiers, including MLP, SVM, and RF were trained on the labeled dataset to spot malicious accounts.

Tajalizadeh and Boostani [68] introduced an innovative clustering framework that used INB instead of Euclidean distance and applied DenStream to improve the performance of assigning new samples to the clusters more accurately. Sinha, et al. [69] gathered Twitter data and focused on studying the ecosystem of spammers, i.e., reportees, as well as reporters who reported spammers by employing DT, KNN, RF, and graph analysis of reporter-reportee clusters by using Gephi application to prevent benign Twitter users from being misclassified. In Karakaşlı, et al. [75], authors applied clustering and classification algorithms on different features to classify spammers on Twitter.

Murugan and Devi [73] used DT, PSO, and GA (PSG-DT) along with various types of features to detect a sequence of spam tweets in real-time. Singh, et al. [71] developed a novel framework to identify the true identity of Twitter users in real-time by employing content, user, and graph-based features. Many researchers, [84], [66], [74], [85], [7], [86], [78], [67], [81], and [83], adopted hybrid techniques by applying a combination of various types of features to identify spammer profiles.

The papers related to the classification of hybrid analysis approaches, along with their main ideas, merits, demerits, and evaluation parameters are tabulated in Table 10.

## 5 ANALYSIS OF RESULTS

In this section, we analyze the results of this SLR according to the defined protocol in Section 3. An overview of the studied papers is given in Section 5.1. Further, a comparison of them is presented in Section 5.2 by answering the aforementioned research questions stated in Table 3.

### 5.1 Overview of the Selected Studies

To investigate the state-of-the-art on Twitter spam detection approaches, the following complementary questions are specified:

- Which publishers have published the highest number of papers on Twitter spam detection?
- Where are the active research communities, and their research focuses on Twitter spam detection techniques?
- How was the distribution of studies per publication channel in Twitter spam detection?

In this subsection, the distribution of 55 studied papers—categorized by publishers— and the year of publication over time are depicted in Fig. 5. It indicates that IEEE, Springer, ACM, and ScienceDirect have presented the highest number of papers on this topic. Additionally, Wiley and Taylor&Francis have published the least number of papers.

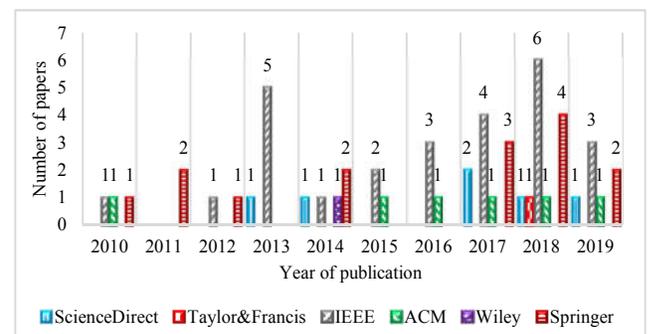

Fig. 5. The number of selected papers classified by publishers and years

As shown in Table 11, among 55 studied papers, 13 papers were published in TCSS, IEEE Access, TIFS, ICACT, and CCS. The distribution of articles per publication channel is depicted in Table 11. Table 12 shows the distribution of active research communities in the studied papers by considering the author's affiliations after the *Final selection* in the *Conducting phase*. Active communities that proposed at least two included studies are listed along with their research focus in Table 12.

TABLE 10
REVIEWING AND COMPARING HYBRID ANALYSIS APPROACHES

| Ref. | Main ideas | Merits | Demerits | Evaluation Parameters |
|---|---|---|---|---|
| [79] | Detecting Twitter spam based on novel content-based and graph-based features | • High F-measure | • Low precision | Precision, recall, F-measure |
| [72] | Using classification for Twitter spam detection | • High precision<br>• High F-measure | • Not tested on a large dataset (low scalability) | Precision, recall, F-measure |
| [62] | Applying hybrid features and classification algorithms for Twitter spam detection | • High F-measure on continuous imbalanced dataset | • Low F-measure of random imbalanced datasets | F-measure, recall, FNR |
| [63] | Proposing statistical features for real-time spam detection | • High accuracy | • High running time | Accuracy, F-measure, detection rate |
| [64] | Presenting effective features for detecting spam accounts | • High recall<br>• High precision | • Not tested on imbalanced datasets | Precision, recall |
| [80] | Detecting spam campaigns on Twitter | • High scalability<br>• High precision<br>• High recall | • Not tested on the text of tweets to improve similarity estimation | Precision, recall, F-measure |
| [65] | Proposing an asymmetric self-learning (ASL) method for spam detection | • Improving detection rate and F-measure | • Not tested on imbalanced datasets<br>• Not detecting spam in real-time | F-measure, detection rate |
| [11] | Proposing an ensemble model for spam detection in an imbalanced dataset | • Improving spam detection rate on imbalanced datasets | • Not tested in the case of considering correlations among features | Precision, F-measure, FPR, recall |
| [88] | Detecting Twitter spam accounts by geographic features | • High F-measure for RF | • Low scalability<br>• Low accuracy | Accuracy, precision, F-measure, recall, root mean square error |
| [70] | Applying statistical features for detecting drifting Twitter spam | • High detection rate<br>• High AUC | • Low scalability | ROC (AUC), accuracy, detection rate |
| [76] | Applying clustering methods for Twitter spam detection | • High recall<br>• High accuracy | • Low F-measure | ROC (AUC), accuracy, precision, recall, F-measure, FPR, specificity |
| [89] | Proposing an automatic classification system applying machine learning to classify spam campaign | • High accuracy | • Not tested on various features | Accuracy, FPR, FNR |
| [77] | Proposing a semi-supervised approach for stream spam detection | • High accuracy | • Applying simple methods due to using limited features | Precision, recall, F-measure |
| [82] | Employing a combination of network science, natural language processing, and ML approaches for predicting the dissemination of tweets and spam tweets propagation | • High scalability<br>• High precision<br>• High recall | • Not considering overlapping among communities | Accuracy, precision, recall, F-measure |
| [90] | Offering an unsupervised method for spammer community detection in Twitter | • High F-measure in detecting spammer communities<br>• High robust | • Not able to detect spammer communities in real-time | Local clustering coefficient, Jaccard similarity coefficient, accuracy, precision, recall, F-measure |
| [87] | Offering a spam detection method based on classification and clustering | • High accuracy for RF classifier | • Low scalability<br>• Not considering the semantic analysis of tweet contents by applying NLP features | Time, precision, recall, F-measure, FPR, FNR, specificity |
| [68] | Presenting a clustering framework for spam detection | • High F-measure<br>• High recall<br>• High purity | • High computational overhead<br>• Not performing distance learning in real-time | Precision, recall, F-measure, purity, computational complexity |
| [69] | Applying data mining techniques for spammer detection | • High accuracy | • Not analyzing the behavior of spammer reporters and reportees | Precision, recall, F-measure |
| [75] | Proposing a dynamic feature selection model for spam detection | • High accuracy | • Low scalability | Accuracy |
| [73] | Proposing a hybrid approach for detecting Twitter stream spam | • High accuracy | • Low detection rate in the imbalanced dataset | Precision, recall, F-measure, FPR |
| [71] | Presenting a hybrid approach to detect different types of spammers by applying graph and content-based features | • High detection rate | • Not updating the classification features due to the change in spammers' behaviors | Accuracy, precision, recall, FPR, FNR, specificity |
| [84] | Offering a hybrid spam detection technique by using content-based and graph-based features | • High accuracy<br>• Low FPR | • Not tested on tweet-level features | Precision, FPR, recall |
| [66] | Employing statistical features for spam detection | • High accuracy of RF and C 5.0 | • Low F-measure on imbalanced dataset | Accuracy, precision, F-measure, scalability, FPR, recall |
| [74] | Applying various numerical features for spam detection | • High ROC (AUC) | • Low scalability<br>• Not tested on imbalanced datasets | ROC (AUC) |
| [85] | Spam profile detection based on classification | • High scalability<br>• High accuracy | • High running time | Accuracy, precision, recall, specificity |
| [7] | Proposing a NN-based approach for spam detection | • High accuracy | • Low F-measure in an imbalanced dataset | Time, ROC (AUC), accuracy, precision, recall, F-measure |
| [86] | Presenting a hybrid method for spam account detection | • High accuracy<br>• High precision<br>• High F-measure | • Not tested on and evaluating imbalanced datasets | Accuracy, precision, recall, F-measure |
| [78] | Proposing a spammer detection system in Twitter | • High accuracy<br>• High F-measure | • Not tested on and not evaluating graph-based features | Time, accuracy, precision, F-measure, FPR, recall |
| [67] | Offering a classification method for drifted Twitter spam | • High accuracy | • High cost | Accuracy, recall, F-measure |
| [81] | Introducing a honeypot-based spammer detection system | • High scalability | • Low recall | ROC (AUC), accuracy, precision, recall, F-measure |
| [83] | Applying ML algorithms for spam detection | • High scalability<br>• High F-measure | • Not tested on and not evaluating the impact of imbalanced datasets on spam detection | Precision, recall, F-measure |

TABLE 11
THE DISTRIBUTION OF THE STUDIES PER PUBLICATION CHANNEL

| Category | Publisher | Publication channel | Count |
|---|---|---|---|
| Journals | IEEE | IEEE Transactions on Computational Social Systems (TCSS) | 3 |
| | | IEEE Access | 3 |
| | | IEEE Transactions on Information Forensics and Security (TIFS) | 2 |
| | Science Direct | Neurocomputing (NEUCOM) | 1 |
| | | Computers & Security (COSE) | 1 |
| | | Expert Systems with Applications (ESWA) | 1 |
| | | Computers & Electrical Engineering (COMPELECENG) | 1 |
| | | Information Sciences (INS) | 1 |
| | Springer | Journal of Supercomputing (SUPE) | 1 |
| | | Wireless Personal Communications (WPC) | 1 |
| | | Wireless Networks (WiNe) | 1 |
| | | Journal of Intelligent Information Systems (JIIS) | 1 |
| | Taylor | Cybernetics and Systems (CybernSyst) | 1 |
| | Wiley | Transactions in GIS (TGIS) | 1 |
| Conferences | IEEE | International Conference on Advanced Communication Technology (ICACT) | 2 |
| | | IEEE International Conference on Collaborative Computing: Networking, Applications and Worksharing (CollaborateCom) | 1 |
| | | International Conference on Computer Engineering and Systems (ICCES) | 1 |
| | | IEEE International Conference on Smart Computing (SMARTCOMP) | 1 |
| | | Latin American Computer Conference (CLEI) | 1 |
| | | International Conference on Social Computing (SocialCom) | 1 |
| | | International conference on security and cryptography (SECRYPT) | 1 |
| | | International conference on communications (ICC) | 1 |
| | | International conference on data mining (ICDM) | 1 |
| | | Conference on Computer Communications Workshops (INFOCOM WKSHPS) | 1 |
| | | International Bhurban Conference on Applied Sciences and Technology (IBCAST) | 1 |
| | | International Conference on Semantics, Knowledge and Grids (SKG) | 1 |
| | | International Conference on Advances in Social Networks Analysis and Mining (ASONAM) | 1 |
| | | International Conference on Advances in Computing, Communications and Informatics (ICACCI) | 1 |
| | | International Conference on Recent Trends in Information Technology (ICRTIT) | 1 |
| | | International Conference on Industrial and Information Systems (ICIIS) | 1 |
| | | International Conference on Innovation and Intelligence for Informatics, Computing, and Technologies (3ICT) | 1 |
| | ACM | ACM Conference on Computer and Communications Security (CCS) | 3 |
| | | International ACM SIGIR Conference on Research and Development in Information Retrieval (SIGIR) | 1 |
| | | Collaboration, electronic messaging, anti-abuse and spam conference (CEAS) | 1 |
| | | World Wide Web Conference (WWW) | 1 |
| | ScienceDirect | Procedia Computer Science (Procedia Comput Sci) | 1 |
| | Springer | International Joint Conference SOCO'13-CISIS'13-ICEUTE'13 | 1 |
| | | International Conference on Security & Privacy (ISEA-ISAP) | 1 |
| | | International Conference on Industrial, Engineering and Other Applications of Applied Intelligent Systems (IEA/AIE) | 1 |
| | | International workshop on recent advances in intrusion detection (RAID) | 1 |
| | | International Conference on Technologies and Applications of Artificial Intelligence (TAAI) | 1 |
| | | International Conference on Autonomic and Trusted Computing (ATC) | 1 |
| | | International Conference on Applied Cryptography and Network Security (ACNS) | 1 |
| | | International Telecommunications Conference (ITelCon) | 1 |
| | | OTM Confederated International Conferences "On the Move to Meaningful Internet Systems" (OTM) | 1 |
| | | the International Conference on Signal, Networks, Computing, and Systems (ICSNCS) | 1 |
| | | International Conference on E-Business and Telecommunications (ICETE) | 1 |

TABLE 12
ACTIVE COMMUNITIES AND THEIR RESEARCH FOCUS

| Affiliation | Ref. | Category |
|---|---|---|
| Deakin University, Australia | [11], [62], [63], [65], [66], [70] | Hybrid analysis |
| Pennsylvania State University, USA | [76], [83], [79] | Hybrid analysis |
| Georgia Institute of Technology Atlanta, USA | [39], [85] | Content analysis, Hybrid analysis |
| Nanyang Technological University, Singapore | [44], [77] | Content analysis, Hybrid analysis |
| Southwest University, China | [62], [65] | Hybrid analysis |
| Birzeit University, Ramallah, Palestine | [51], [52] | Tweet analysis |
| National Institute of Technology Karnataka, India | [42], [90] | Content analysis, Hybrid analysis |

A significant number of studied papers are published by Deakin University in Australia and Pennsylvania State University in the USA. Moreover, among the studied papers, researchers at Georgia Institute of Technology, Atlanta in the USA; Nanyang Technological University in Singapore; Southwest University in China; Birzeit University, Ramallah in Palestine; and National Institute of Technology Karnataka in India have published 2 papers on this topic each. As depicted in Table 12, most of the researchers focused on Hybrid analysis approaches to detect spam on Twitter.

## 5.2 Research Objectives, Techniques, and Evaluation Parameters

The results of this SLR, addressing the answers to the research questions RQ1, RQ2, RQ3, and RQ4 (shown in Table 3), are described in this subsection. Considering the literature, in this SLR, to answer the research questions, the percentage of each item was calculated by applying (1). In this equation, after counting the number of each occurrence, it was divided by the sum of the total number of occurrences, then the answer was multiplied by 100.

$$\text{Percentage of occurrence (i)} = \frac{\text{Number of each occurrence}}{\sum \text{Number of all occurrence}} * 100 \quad (1)$$

Considering RQ1, due to the proliferation of the Internet and the expansion of electronic and mobile devices around the world, sharing information on online social networks has increased. Among these networks, Twitter has become the target of spammers due to its growing popularity and the ease of use. Since spammers are aware of the available spam detection techniques on Twitter platform, spammers are constantly changing the way of attack, so it is difficult to detect such malicious users on Twitter.

Spammers take advantage of Twitter to pretend to be a legitimate user and try to propagate malicious tweets to make money. Therefore, making a secure Twitter and keeping the privacy of users' information is the essential topic being explored by different researchers. Thus, this SLR will be very useful for researchers to review the works that have been done in this area quickly.

Considering RQ2, based on the reviewed papers discussed in Section 4, Twitter spam detection techniques are categorized in five main approaches: Content analysis ap-proaches, user analysis approaches, tweet analysis approaches, network analysis approaches, and hybrid analysis approaches. It can be seen in Fig. 6 that the highest percentage of studies is conducted by employing hybrid analysis approaches with 56%, while content analysis approaches have 15%. Network analysis approaches have been applied 11% by researchers. User approaches and tweet analysis approaches have been applied 9% of all types of used techniques each. As shown in Fig. 7, RF, SVM, NB, and KNN were the most applied algorithms and classifiers in the studied papers.

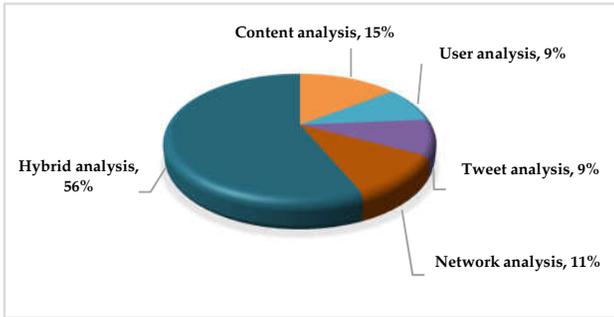

Fig. 6. The percentage of Twitter spam detection approaches in the selected papers

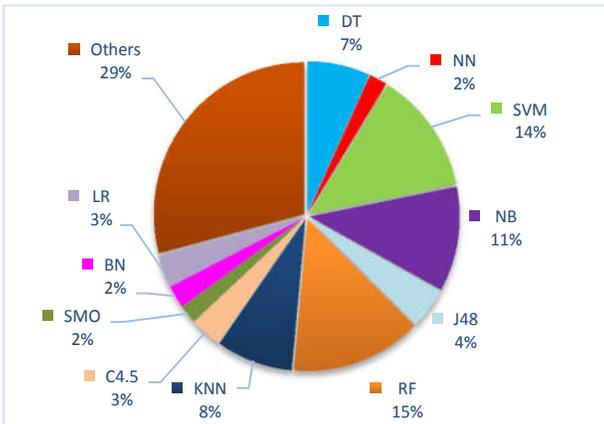

Fig. 7.The percentage of applied algorithms on Twitter spam detection approaches

According to the obtained information from Tables 6, 7, 8, 9, and 10, a comparison of similarities and differences among the current approaches in Twitter spam detection is provided in Table 13.

In content analysis approaches, researchers have paid more attention to factors such as accuracy and scalability. User analysis approaches have tried to improve the accuracy and F-measure. Tweet analysis approaches have focused on improving recall. In network analysis approaches, factors such as detection rate, accuracy, and F-measure are regarded more. Table 13 also illustrates that the accuracy, F-measure, recall, and precision have improved in hybrid analysis approaches. On the other hand, high running and response time, high cost, and low scalability are the main disadvantages of most approaches. Moreover, hybrid analysis approaches have not investigated the impact of the various distribution of classes in

TABLE 13
MAIN ADVANTAGES AND DISADVANTAGES OF THE DISCUSSED TAXONOMY

| Approach | Merits | Demerits |
|---|---|---|
| Content analysis approaches | • Better accuracy<br>• Better scalability | • Unacceptable response time |
| User analysis approaches | • Better accuracy<br>• Better F-measure | • Unacceptable running time |
| Tweet analysis approaches | • Better recall | • Unacceptable scalability |
| Network analysis approaches | • Better accuracy<br>• Better detection rate<br>• Better F-measure | • Unacceptable scalability<br>• Unacceptable cost |
| Hybrid analysis approaches | • Better accuracy<br>• Better F-measure<br>• Better recall<br>• Better precision | • Not evaluated on the various distribution of classes<br>• Unacceptable F-measure in imbalanced datasets<br>• Unacceptable cost<br>• Unacceptable running time |

Twitter spam detection, and in the case of testing on imbalanced datasets, it had low F-measure. Thus, since the real datasets are imbalanced, researchers have to improve their proposed techniques on imbalanced datasets.

Considering RQ3, researchers have employed various evaluation parameters. The definition of evaluation parameters applied in the studied papers is tabulated in Table 14. After applying (1) that obtains the percentage of each specific parameters regarding the other ones, the parameters and their percentages are illustrated in Fig. 8. As shown in Fig. 8, the highest percentage of evaluation parameters in Twitter spam detection is related to recall parameter with 23%, and F-measure comes next with 18%, precision with 17%, and accuracy 14%. A comparative investigation of the evaluation parameters of the chosen papers in each approach is provided in this section in Fig. 9.

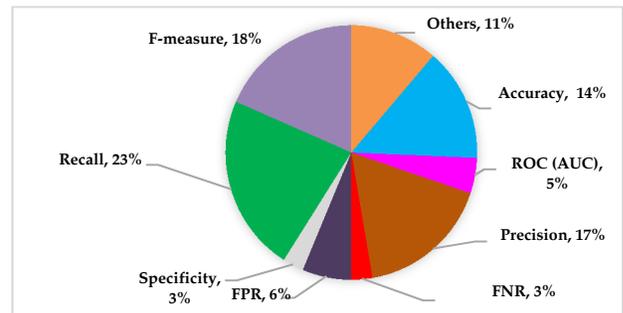

Fig. 8. The percentage of evaluation parameters in twitter spam detection

Fig. 9 indicates that, in content analysis approaches, researchers focused on recall, precision, accuracy, F-measure, and ROC. Moreover, in user analysis approaches, F-measure, recall, accuracy, precision, and FPR are the crucial ones. Recall, F-measure, precision, accuracy, and specificity are also significant in tweet analysis approaches.

The papers with network analysis approaches have optimized recall, F-measure, precision, accuracy, and ROC. To say more, in hybrid analysis approaches, recall, F-measure, precision, accuracy, and FPR are essential. So, the results showed that all of the approaches made attempts to enhance recall, F-measure, precision, and accuracy.

TABLE 14
LIST OF EVALUATION PARAMETERS AND THEIR DESCRIPTION

| Parameter name | Description |
|---|---|
| Confusion matrix | By considering 1 for spam and 0 for non-spam, four possible outputs of the confusion matrix are the 2*2 elements as below: TP: If spam tweets are correctly classified, it is called a True Positive value. TN: If non-spam tweets are correctly classified, it is called a True Negative value. FP: If non-spam tweets are classified as spam, it is considered False Positive. FN: If spam tweets classified incorrectly as non-spam tweets, it is called False Negative. |
| Accuracy | It is the proportion of correctly predicted observations over the total number of observations. It can be calculated by this formula: $Accuracy = \frac{TP+TN}{TP+FP+FN+TN}$ |
| Precision | It is the ratio of positive observations predicted correctly over the total observations predicted positive. It can be measured by this formula: $Precision = \frac{TP}{TP+FP}$ |
| Recall or Sensitivity or TPR | It is the proportion of positive observations that are classified correctly over all positive cases in the actual class of the dataset. It is measured by this formula: $Recall = \frac{TP}{TP+FN}$ |
| F-measure | It is the harmonic mean of precision and recall value. It is measured by this formula: $F-mesure = 2 * \frac{precision*recall}{precision+recall}$ |
| ROC (AUC) | It is the plot between recall (sensitivity) on the Y-axis and (1-specificity) on the X-axis and the area under this curve referred to as AUC, which indicates the ability of a model to distinguish positive and negative samples. |
| Specificity or TNR | It is the proportion of actual negative observations that are classified correctly. It is measured by $Specificity = \frac{TN}{TN+FP}$ |
| FPR | The rate of classifying non-spam tweets as spam tweets. It can be measured by this formula: $FPR = \frac{FP}{FP+TN}$ |
| FNR | The rate of classifying spam tweets as non-spam tweets. It can be measured by this formula: $FNR = \frac{FN}{FN+TP}$ |
| Time | It can be calculated by the total time that is taken to build a model on the training dataset plus the time to make a prediction based on the testing dataset. |
| Transitivity | It calculates the tendency of edges to form triangles in a graph. It is also called the clustering coefficient, and it measures the probability of connecting adjacent vertices of a node. |
| Degree variance | In a graph with n nodes and m edges, the variance of node degrees is calculated by this formula: $Degree\ variance = \frac{2m(n^2-n-2m)}{n^3+n^2}$ |
| Local clustering co-efficient | It quantifies how close the neighbors of a node are to make a clique (complete graph) or fraction of pairs of the friends of a node that are friends with one another. |
| Jaccard similarity coefficient | It is the ratio of the number of members shared between two datasets to the number of union members in two datasets. $Jaccard\ similarity\ coefficient = \frac{|A \cap B|}{|A \cup B|}$ |
| Cosine similarity distance | It is applied to capture similarities between various documents/vectors. The degree of angle between two vectors/documents is measured (the frequency of a term in various documents collected) |
| Root mean square error (RMSE) | It measures the square root of the mean of squared differences between predictions and actual observations. It is measured by $RMSE = \sqrt[2]{\frac{1}{n}\sum_{j=1}^{n}(y_j - \hat{y}_j)^2}$ |
| Scalability | It is the capability of a model to cope well under the increasing volume of datasets or high workload. |
| Detection rate | It is the ratio of positive observations predicted correctly over the total predictions. It is measured by $Detection\ rate = \frac{TP}{FN+TN+FP+TP}$ |
| Purity | This is a metric for qualifying clustering. Purity examines to what extent a cluster has a single class. It is calculated by the summation of the number of elements from the most dominant class in each cluster and then divides the total sum to the number of all elements. It is measured by $Purity = \frac{1}{N}\sum_{i=1}^{K} max_j (c_i \cap t_j)$ |
| Computational complexity | It refers to time complexity or space complexity. Time complexity denotes the speed of a classifier for the input dataset, while space complexity refers to the amount of memory an algorithm or classifier needs concerning the size of the input dataset. |
| Betweenness centrality | It calculates the extent to which a node lies on the shortest paths that pass through a pair of vertices. |
| Maximum cluster diameter | It calculates the maximum distance between pairs of elements within the same cluster. |

However, FPR, ROC, specificity, and FNR are somewhat neglected in these approaches.

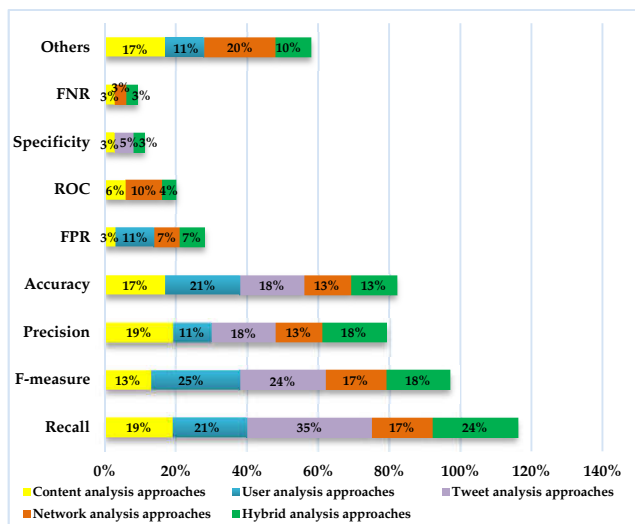

Fig. 9. The number of evaluation parameters in each approach of the selected papers

Considering RQ4, in Fig. 10, the percentage of evaluation tools in reviewed papers in Twitter spam detection is demonstrated statistically. Weka has 28% of usage, Python and R programming language have 15% and 5%, respectively.

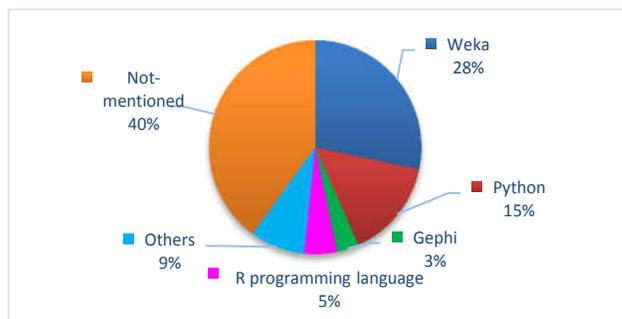

Fig. 10. The percentage of evaluation tools in Twitter spam detection.

Fig. 11 depicts the evaluation techniques applied in the papers reviewed in Section 4. There were four types of evaluation techniques in the reviewed papers: Simulation, prototype, data sets, and design. As shown in Fig. 11, most evaluation techniques are related to data sets, with 74%, 15% of assessments were related to simulation, and 9% of them were prototype, while 2% of evaluation techniques were related to design.

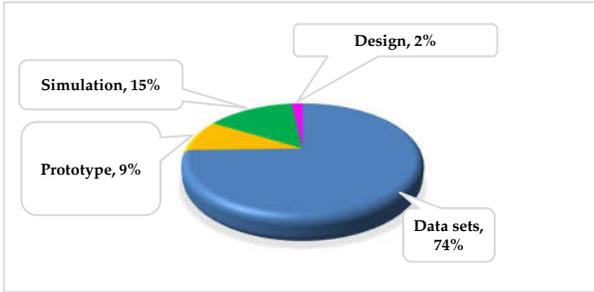

Fig. 11. The percentage of evaluation techniques in the selected papers

Fig. 12, displays the repetition of evaluation techniques in each analysis approach. Since datasets were the most widely used in innovations of spam detection approaches, the comparative results illustrate that, they have the highest percentages in most assessments of each approach.

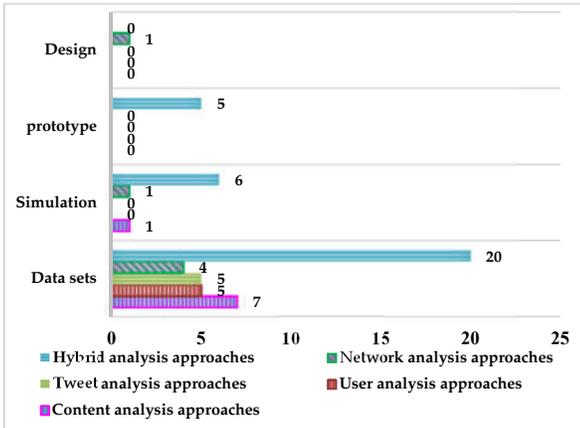

Fig. 12. Repetition of types of evaluations in each approach in the selected papers

## 6 OPEN ISSUES AND FUTURE DIRECTIONS

By analyzing all the reviewed papers devoted to the topic of spam detection, we identify some significant issues that have not been investigated. Hence, considering RQ5, this section discusses the open problems and challenges in this topic in detail.

- *Real-time Twitter data collection:* Unlike face-to-face interactions, social media activities, can be captured. Although Twitter made it possible for researchers to extract near real-time tweets through Twitter's Streaming API in a sampled form, due to the keyword-based algorithms, the real-time dataset collection cannot be reproduced for a defined query, and the tweets sampling method is unclear. So, it is challenging to capture a Twitter dataset, especially real-time or unbiased ones.

- *Labeling dataset.* After collecting tweets, in supervised approaches, the data should be labeled to spam or non-spam by applying one of the manual inspection or blacklist filtering. While manual inspection is very accurate, it can only label a small amount of training data, so it is a time-consuming task and requires a large number of people and resources and needs a great effort. Leveraging crowdsourcing techniques, such as Amazon Turk, is a common approach in social science to mitigate this problem, it yet is insufficient. Similarly, employing blacklist services such as GSB allows researchers to label more tweets, but it is not very accurate and increases the error of classifier evaluations. Additionally, to maintain the effectiveness of spam detection, due to the changes in the pattern of sending spam by spammers, labeling should be done repeatedly, which is so costly.

- *Spam drift problem.* Observing the extracted Twitter data indicates that the characteristics of spam tweets are changing over time, which refers to "spam drift". This is owing to the fact that spammers are constantly changing their attack patterns to conceal from the radar of spam detectors. Unfortunately, ML algorithms are not updated with the "varied" spam tweets; thus, the performance of classifiers is reduced dramatically. Few articles offer solutions to solve the problem of "spam drift", and this is still one of the challenges for researchers.

- *Class imbalance problem.* Some of the studies have ignored the imbalanced problem in real-world Twitter dataset, and they construct a balanced dataset of positive and negative samples. The uneven distribution of spam and non-spam classes significantly reduces the classification performance, due to intrinsic nature of ML algorithms that usually perform better for the majority (non-spam) class and ignore the minority (spam) class. Although some researchers have proposed methods to enhance spam detection in imbalanced datasets, it yet needs more research and is a significant challenge.

- *Low scalability and large-scale Twitter dataset.* Like other ML detection schemes, large-scale ground truth is also a critical source for understanding and developing advanced classification and clustering algorithms, as well as for presenting critical training and testing data for such algorithms. Some of the proposed methods have been trained and evaluated on a small dataset, and are not scalable. Apart from the classification algorithms, numerous articles reviewed in the literature apply clustering techniques to identify spam in bulk. Although clustering techniques do not require pre-labeled data, they suffer from the drawback of the ever-increasing size of social networks, which is a main open challenge of identifying spam campaigns.

- *Resampling the training dataset and tuning hyperparameters.* Apart from applying various evaluation parameters, tuning the hyperparameters of the model along with resampling the dataset to create a balanced dataset, were somewhat neglected. So, utilizing various metaheuristic algorithms for optimizing hyperparameters of classifiers or for selecting the most suitable features are further directions in this domain.

- *Length of tweets.* Twitter, initially allowing users to

send up to 140- character tweets, increased the tweet length to 280 characters in November 2017. Another research direction is to examine the effect of increasing the length of tweets on spam detection approaches and spammer activities.

- *Graph-based features.* Some papers extract graph-based features to distinguish legitimate users from fake ones. In comparison with other kinds of features, graph-based features such as clustering coefficient and betweenness centrality are robust and perform better at detecting abnormal behaviors on Twitter, but extraction of these features are expensive and time-consuming. Because the spam accounts behavior like their following or tweeting behavior can be controlled or changed by spammers, graph-based features or neighbor-based features are hard to evade.
- *Streaming data analysis:* In addition to collecting real-time Twitter datasets, analyzing Twitter data streams is a significant challenge which faces three major challenges, including volume, velocity, and volatility. Even if the high volume of real-time tweets can be collected, the vast amount of data that is coming constantly with high speed, generating varying features called feature drift, makes the problem of streaming data analysis an open challenge.
- *Streaming data preprocessing:* Since labelling dataset, identifying optimal models and classifiers in a supervised approach require effort, it is not appropriate for real-time processing. However, this vast amount of data stream can be analyzed offline by unsupervised or semi-supervised approaches such as clustering algorithms, since clustering algorithms do not require labelling and processing, but the existence of new techniques and algorithms to analyze data streams with low memory and time consumption is still a significant challenge. Moreover, since data processing is an important procedure in data analysis, and streaming data, which comes continuously, may be noisy, incomplete, redundant, and even contain missing values, due to changing nature of streaming tweets, manual preprocessing is not practical and automated preprocessing approaches is needed for feeding the model with the preprocessed data.
- *Non-English tweets.* Analyzing and detecting spams of non-English tweets is another future research direction. Applying various languages in social networks leads to unsolved challenges. Most of the automatic detection tools are usually built for English language and there is a lack of relevant tools and sufficient resources for other languages such as Chinese, Hindi, French, Spanish, and so on. Different languages have various features, prefixes, suffixes, and syntactic and semantic structures. Hence, it is necessary to introduce new tools and extend novel approaches to analyze and handle the linguistic diversity and non-English features on real-time tweets.

## 7 THREATS TO VALIDITY

This SLR presents a taxonomy and a comparison of Twitter spam detection approaches. These types of review papers usually have limitations [13], but the results of SLR are mainly reliable [91]. The major limitations and threats to the validity of this SLR are discussed in this section.

In order to cover as much as possible relevant papers in this topic and prevent any bias, we use a combination of various terms. In this respect, to retrieve literature from seven academic databases and ensure unbiased ones, a review protocol was defined. After the *initial selection step*, that retrieved 356 papers, the inclusion/exclusion criteria were employed to filter the most relevant 55 papers to answer predefined questions. We only consider academic journals and conferences, so one of the limitations is ignoring editorial papers, white papers, papers less than six pages, non-English papers, book chapters, theses, and review papers. Hence, selecting all the relevant papers is not guaranteed.

Another limitation is defining five research questions to find their answers by investigating related papers. Other researchers may define some other research questions. Moreover, Twitter spam detection approaches were categorized into five groups, but they can be classified otherwise. In a nutshell, by defining a review protocol, follow a systematic manner, and the involvement of various researchers, the validity of this SLR is high.

## 8 CONCLUSION

In this paper, a comprehensive SLR has been conducted covering 55 most relevant papers out of 356 papers published between 2010 and 2020. Each paper has been investigated along with describing the research methodology, statistical analysis of the applied approaches, tools, evaluation parameters, and evaluation techniques. With respect to RQ3, it was found that the most widely considered evaluation parameters were recall (23%), F-measure (18%), precision (17%), and accuracy (14%), but the FPR, ROC, FNR, and specificity were somewhat ignored. Based on RQ4, it was clear that in the studied papers, Weka had the highest percentage of usage of all evaluation tools. Additionally, a taxonomy was presented to draw a lucid picture of Twitter spam detection approaches. We classified Twitter spam detection approaches based on feature analysis into five categories: Content analysis approaches (15%), user analysis approaches (9%), tweet analysis approaches (9%), network analysis approaches (11%), and hybrid analysis approaches (56%). Finally, we highlighted open issues, challenges, and future directions.

**Sepideh Bazzaz Abkenar** received her B.S. in Computer Engineering, Software Engineering, from Tabarestan Higher Education Institution (TU), Chaloos, Iran, in 2008; She is a M.S. student in Information Technology, Computer Networks, in Department of Computer Engineering, Science and Research Branch, Islamic Azad University, Tehran, Iran. Her research interests include big data analytics, social networks, and machine learning.

**Mostafa Haghi Kashani** received his BS in Computer Engineering from Kashan Branch of IAU, Iran, in 1999 and the MS in Computer Engineering from South Tehran Branch of IAU, Iran in 2002. He is currently a full-time PhD Candidate in Computer Engineering-Software Systems at Science and Research Branch of IAU, Tehran, Iran. He is a researcher and lecturer in the Department of Computer Engineering at the IAU University. He has authored/co-authored several papers in technical journals and conferences. His research interests include distributed systems, fog computing, social networks and evolutionary computing. He has acted as a reviewer in several international journals, including the Journal of Supercomputing (Springer), the International Journal of Communication Systems (Wiley), and the Journal of Big Data (Springer).

**Mohammad Akbari** received the PhD degree from the School of Integrative Sciences and Engineering, National University of Singapore, 2016. He is currently an assistant professor in the Department of Computer Science, Amirkabir University of Technology. His research interests span mainly in AI and intelligent system including reinforcement learning, deep learning, data mining and machine learning using large-scale datasets, with an emphasis on their applications in health informatics, social informatics, information retrieval, recommendation system and smart cities. His research has been published in several major academic venues, including SIGIR, WSDM, ICMR, etc. One of his paper has been selected as the best paper in IJCAI 2016. He has served as senior program committees for ACM MM and ICWSM and program committee for several major conferences such as KDD, ACL and ECIR, and reviewed for multiple journals, including the IEEE Transactions on Knowledge and Data Engineering and the ACM Transactions on Information Systems.

**Ebrahim Mahdipour** received the B.S. degree in computer engineering, specialized in hardware engineering, from the Dezful Branch, Islamic Azad University, Dezful, Iran, in 2003, the M.S. degree in computer engineering, specialized in computer architecture, and the Ph.D. degree in computer engineering, specialized in computer architecture, from the Science and Research Branch, Islamic Azad University, Tehran, Iran, in 2006 and 2012. He is the founding director of the Cyber Security Research Center, and he is currently an Assistance Professor with the Department of Computer Engineering, Science and Research Branch, Islamic Azad University. His research interests include cyber security, blockchain and big data.